\documentclass[apjl]{emulateapj}
\usepackage[colorlinks=true,citecolor=blue,urlcolor=red,linkcolor=blue]{hyperref}

\def\HII{\hbox{H\,{\sc ii}}}

\def\CII{\hbox{[C\,{\sc ii}]157.7$\,\mu$m}}
\def\CIIno{\hbox{[C\,{\sc ii}]}}

\def\OIIIno{\hbox{[O\,{\sc iii}]}}

\def\NIIno{\hbox{[N\,{\sc ii}]}}
\def\NIIIno{\hbox{[N\,{\sc iii}]}}

\def\H2{\hbox{H$_{2}$}}

\def\Lsun{\hbox{$L_\odot$}}

\def\LIR{\hbox{$L_{\rm IR}$}}

\def\SIR{\hbox{$\Sigma_{\rm IR}$}}

\def\nH{\hbox{$n_{\rm H}$}}
\def\G0{\hbox{$G_{\rm 0}$}}

\def\c-m{\hbox{cm$^{-1}$}}
\def\c-mm{\hbox{cm$^{-2}$}}
\def\c-mmm{\hbox{cm$^{-3}$}}

\def\k-pcc{\hbox{kpc$^{-2}$}}

\def\Tdust{\hbox{$T_{\rm dust}$}}

\shorttitle{Extended \CIIno\, Emission in Local LIRGs}
\shortauthors{D\'iaz-Santos et al.}

\begin{document}

\title{Extended \CIIno\, Emission in Local Luminous Infrared Galaxies}


\author{T.~D\'{\i}az-Santos\altaffilmark{1,*},
L.~Armus\altaffilmark{1},
V.~Charmandaris\altaffilmark{2,3,4},
G.~Stacey\altaffilmark{5},
E.~J.~Murphy\altaffilmark{6},
S.~Haan\altaffilmark{7},
S.~Stierwalt\altaffilmark{8},
S.~Malhotra\altaffilmark{9},
P.~Appleton\altaffilmark{10},
H.~Inami\altaffilmark{11},
G.~E.~Magdis\altaffilmark{12},
D.~Elbaz\altaffilmark{13}
A.~S.~Evans\altaffilmark{8,14},
J.~M.~Mazzarella\altaffilmark{15},
J.~A.~Surace\altaffilmark{1},
P.~P.~van~der~Werf\altaffilmark{16},
C.~K.~Xu\altaffilmark{15},
N.~Lu\altaffilmark{15},
R.~Meijerink\altaffilmark{16,17},
J.~H.~Howell\altaffilmark{15},
A.~O.~Petric\altaffilmark{18,19},
S.~Veilleux\altaffilmark{20,21},
\& D.~B.~Sanders\altaffilmark{22}
%
%B.~Schulz\altaffilmark{12,17}
%S.~Lord\altaffilmark{16}
%C.~Bridge\altaffilmark{12},
%B.~H.~P.~Chan\altaffilmark{12},
%D.~T.~Frayer\altaffilmark{19},
%K.~Iwasawa\altaffilmark{20},
%and E.~Sturm\altaffilmark{22}
}

\altaffiltext{*}{Contact email: tanio@ipac.caltech.edu}
\affil{$^{1}$Spitzer Science Center, California Institute of Technology, MS 220-6, Pasadena, 91125, CA}
\affil{$^{2}$Department of Physics, University of Crete, GR-71003, Heraklion, Greece}
\affil{$^{3}$Chercheur Associ\'e, Observatoire de Paris, F-75014, Paris, France}
\affil{$^{4}$Institute for Astronomy, Astrophysics, Space Applications \& Remote Sensing, National Observatory of Athens, GR-15236, Athens, Greece}
\affil{$^{5}$Department of Astronomy, Cornell University, Ithaca, NY 14853, USA}
\affil{$^{6}$Observatories of the Carnegie Institution for Science, 813 Santa Barbara Street, Pasadena, CA 91101, USA}
\affil{$^{7}$CSIRO Astronomy and Space Science, Marsfield NSW, 2122, Australia}
\affil{$^{8}$Department of Astronomy, University of Virginia, P.O. Box 400325, Charlottesville, VA 22904}
\affil{$^{9}$School of Earth and Space Exploration, Arizona State University, Tempe, AZ 85287, USA}
\affil{$^{10}$NASA Herschel Science Center, IPAC, California Institute of Technology, MS 100-22, Cech, Pasadena, CA 91125}
\affil{$^{11}$National Optical Astronomy Observatory, 950 N. Cherry Ave., Tucson, AZ 85719, USA}
\affil{$^{12}$Department of Physics, University of Oxford, Keble Road, Oxford OX1 3RH, UK}
\affil{$^{13}$Laboratoire AIM-Paris-Saclay, CEA/DSM/Irfu, CNRS, Universite Paris Diderot, Saclay, 91191 Gif-sur-Yvette, France}
\affil{$^{14}$National Radio Astronomy Observatory, 520 Edgemont Road, Charlottesville, VA 22903}
\affil{$^{15}$Infrared Processing and Analysis Center, MS 100-22, California Institute of Technology, Pasadena, CA 91125}
\affil{$^{16}$Leiden Observatory, Leiden University, P.O. Box 9513, NL-2300 RA Leiden, The Netherlands}
\affil{$^{17}$Kapteyn Astronomical Institute, University of Groningen, P.O. Box 800, NL-9700 AV Groningen, The Netherlands}
\affil{$^{18}$Gemini Observatory, 670 N. Aohoku Place, Hilo, HI 96720, USA}
\affil{$^{19}$Astronomy Department, California Institute of Technology, Pasadena, CA 91125, USA}
\affil{$^{20}$Joint Space-Science Institute, University of Maryland, College Park, MD 20742, USA}
\affil{$^{21}$Department of Astronomy, University of Maryland, College Park, MD 20742, USA}
\affil{$^{22}$Institute for Astronomy, University of Hawaii, 2680 Woodlawn Drive, Honolulu, HI 96822}
%\affil{$^{19}$National Radio Astronomy Observatory, P.O. Box 2, Green Bank, WV 24944, USA}
%\affil{$^{20}$ICREA and Institut de Cincies del Cosmos (ICC), Universitat de Barcelona (IEEC-UB), Marti i Franques 1, 08028 Barcelona, Spain}
%\affil{$^{21}$Caltech Optical Observatories, Division of Physics, Mathematics and Astronomy, MS 301-17, California Institute of Technology, Pasadena, CA 91125, USA}
%\affil{$^{22}$Max-Planck-Institut f\"{u}r extraterrestrische Physik, Postfach 1312, D-85741 Garching, Germany}

\begin{abstract}

We present \textit{Herschel}/PACS observations of extended \CII\, line emission detected on $\sim\,1-10\,$kpc scales in 60 local luminous infrared galaxies (LIRGs) from the Great Observatories All-sky LIRG Survey (GOALS). We find that most of the extra-nuclear emission show \CIIno/FIR\, ratios $\geq\,4\,\times\,10^{-3}$, larger than the mean ratio seen in the nuclei, and similar to those found in the extended disks of normal star-forming galaxies and the diffuse inter-stellar medium (ISM) of our Galaxy. The \CIIno\, "deficits" found in the most luminous local LIRGs are therefore restricted to their nuclei. There is a trend for LIRGs with warmer nuclei to show larger differences between their nuclear and extra-nuclear \CIIno/FIR\, ratios.
% This arises from the fact that the average dust temperature (\Tdust) in the extended regions remains rather cold regardless of the nuclear FIR color.
%The difference between the extra-nuclear and nuclear \Tdust\, is clearly correlated with the excess of \CIIno\, deficit seen in the nuclei with respect to that observed in the extended emission.
%This is the first time that these differences have been shown to be present in a representative sample local LIRGs.
We find an anti-correlation between \CIIno/FIR\, and the luminosity surface density, \SIR, for the extended emission in the spatially-resolved galaxies. However, there is an offset between this trend and that found for the LIRG nuclei. We use this offset to derive a beam filling-factor for the star-forming regions within the LIRG disks of $\sim\,6\,\%$ relative to their nuclei. We confront the observed trend to photo-dissociation region (PDR) models and find that the slope of the correlation is much shallower than the model predictions. Finally, we compare the correlation found between \CIIno/FIR\, and \SIR\, with measurements of high-redshift starbursting IR-luminous galaxies.
% We propose that the correlation between \CIIno/FIR\, and \SIR\, found for the LIRG nuclei can be used to derive the size of high-redshift star-bursting IR-luminous galaxies for which an unresolved \CIIno/FIR\, measurement is available.
%We find that, after correcting for the filling-factor, the nuclei as well as the extended star formation in all LIRGs, regardless of their \LIR, follow a extremely tight correlation between the \SIR\, and the luminosity surface density of the \CIIno\, line, \SCII, expanding 3 orders of magnitude in \SIR, with $\SIR\,=\SCII^{1.35\pm0.03}$. This suggests that both the \LIR\, and \LCIIno\, are (1) linked through the internal physics of the star-forming regions from which they most likely originate, and (2) their relation is independent of the macroscopic properties of galaxies. This scaling law also implies that, if the dust emits as a volume surrounding the young stars, then the \CIIno\, emission is proportional to $r^{2.2}$, very close to the geometry of emission from a photo-dissociation region.

\end{abstract}


%________________________________________________________________
\section{Introduction}\label{s:intro}

The far-infrared (FIR) \CII\, fine-structure emission line is the dominant gas coolant of the neutral inter-stellar medium (ISM) in normal star-forming galaxies (\citealt{Malhotra1997}). It mostly arises from photo-dissociation regions (PDR) heated by ultraviolet (UV) radiation emitted by young, massive stars, and can account for up to a few percent of the total IR luminosity of a galaxy (\citealt{Stacey1991}; \citealt{Helou2001}). However, the efficiency of the ISM cooling through this emission line is greatly reduced in systems with enhanced radiation field intensity and/or hardness caused by very intense nuclear starbursts
%, which diminishes the overall energy put into gas heating likely due to the competition of dust for UV photons within the \HII\, regions 
(\citealt{Abel2009}; \citealt{Stacey2010}; \citealt{Gracia-Carpio2011}; \citealt{DS2013})
%, and the build up of grain charge for high UV fields which lowers the photo-electric heating efficiency (\citealt{Stacey2010}). 
Indeed, compared with normal, quiescently star-forming galaxies, the relative cooling via the \CIIno\, line with respect to that carried by dust, as measured by the \CIIno/FIR\, ratio\footnote{FIR is the flux emitted within the $42.5-122.5\,\mu$m wavelength range as originally defined in \cite{Helou1988}.}, decreases by more than an order of magnitude from $\sim\,10^{-2}$ in starbursting galaxies with warmer average nuclear dust temperatures and more compact star formation, such as in local luminous infrared galaxies (LIRGs; $\LIR\,\geq\,10^{11-12}\,\Lsun$) (\citealt{DS2013}). In addition, in those sources where an active galactic nucleus (AGN) may contribute to the FIR emission of the galaxy, the \CIIno/FIR\, ratio can drop even further (i.e., there is a larger \CIIno\, "deficit"), up to two orders of magnitude (\citealt{DS2013}; \citealt{Farrah2013}).

In our Galaxy and the Magellanic Clouds, \CIIno\, emission arising from star-forming complexes and the diffuse ISM have been studied in great detail (e.g., \citealt{Wright1991}; \citealt{Shibai1991}; \citealt{Stacey1993}; \citealt{Bennett1994}; \citealt{Poglitsch1995}; \citealt{Yasuda2008}; \citealt{Pineda2012}; \citealt{Lebouteiller2012}). However, before the launch of the \textit{Herschel Space Observatory} (\citealt{Pilbratt2010}), there was only a limited number of extra-galactic sources for which it was possible to disentangle the \CIIno\, emission arising from spatially-resolved star-forming regions from that originating within their nuclei (e.g., \citealt{Madden1993}; \citealt{Nikola2001}; \citealt{Higdon2003}; \citealt{Kramer2005}; \citealt{RF2006}). Now, the Photodetector Array Camera and Spectrometer (PACS; \citealt{Poglitsch2010}) instrument on board \textit{Herschel} has enabled us to obtain measurements of the FIR continuum and emission lines on physical scales of a few hundred pc in a significant number of nearby galaxies (\citealt{Kennicutt2011}; \citealt{Croxall2012}; \citealt{Beirao2012}; \citealt{Kramer2013}), with which we are able to derive the properties (\G0, \nH) of the different phases of the ISM in extra-galactic star-forming regions (e.g., \citealt{Mookerjea2011}; \citealt{Contursi2013}).

Studying the extra-nuclear \CIIno\, in nearby LIRGs is critical not only for improving our understanding of the physical properties of the ISM in star-forming galaxies, but also for establishing local benchmarks with which to compare measurements of this key emission line in distant IR-luminous galaxies. In fact, observations with PdBI and ALMA are already starting to find high-redshift galaxies in which \CIIno\, is detected at distances several kpc away from their nuclei (\citealt{Gallerani2012}; \citealt{Swinbank2012}; \citealt{Riechers2013}).
%Observations with PdBI and ALMA are already starting to detect \CIIno\, at distances of several kpc away from the nuclei in a number of high-redshift galaxies (\citealt{Gallerani2012}; \citealt{Swinbank2012}; \citealt{Riechers2013}). Therefore, studying the extra-nuclear \CIIno\, emission of their local counterparts is critical not only for our understanding of the physical properties of the ISM in nearby LIRGs, but also to establish local benchmarks with which to compare spatially-resolved measurements of this key line in distant IR-luminous galaxies identified in high-$z$ surveys now and during the following years.

This letter is a follow-up to our work in \cite{DS2013} (hereafter DS13), where we presented the results for our \textit{Herschel}/PACS \CIIno\ survey regarding the nuclear emission of LIRGs. Here we study the extended \CIIno\, line emission detected for the first time in a large sample of nearby LIRGs.
%The data is analyzed within the context laid out in our first paper regarding the nuclear \CIIno\, measurements of our galaxy sample (DS13).
In Section~2 we briefly introduce the sample, observations and analysis of the data. In Section~3 we present our results, and in Section~4 we summarize and discuss them.

\section{Sample and Observations}\label{s:irsobs}

\subsection{The GOALS sample}\label{ss:sample}

The Great Observatories All-sky LIRG Survey (GOALS; \citealt{Armus2009}) encompasses the complete luminosity-limited sample of the 202 LIRGs and ULIRGs
% ($\LIR\,\geq\,10^{12}$) 
contained in the \textit{IRAS} Revised Bright Galaxy Sample (RBGS; Sanders et al. 2003) which, in turn, is a complete flux-limited sample of 629 galaxies having \textit{IRAS} $S_{60\,\mu m}\,>\,5.24$\,Jy and Galactic latitudes $|b|\,>\,5\,^\circ$. There are 180 LIRG and 22 ULIRG systems in GOALS and their median redshift is z = 0.0215 (or $\sim\,95.2$\,Mpc).% The description of how the FIR flux and \LIR\, used in this paper were calculated can be found in \citep{DS2013}.

\subsection{Herschel/PACS Observations}\label{ss:pacsobs}

We have obtained \textit{Herschel}/PACS FIR spectroscopic observations of the \CII\, emission line for 200 LIRG systems (241 individual galaxies) in the GOALS sample. Details regarding the processing and analysis of the data are described in DS13. Since the goal of this letter is to analyze the \CIIno\, deficit in the extra-nuclear regions of LIRGs, in addition to the nuclear \CIIno\, and FIR fluxes calculated in DS13, we have also obtained their spatially-integrated fluxes as extracted from a $3\,\times\,3$ spaxel box ($(9.4\arcsec)^2$/spaxel) centered around the nuclear spaxel. The extended emission is thus defined as the integrated flux minus the nuclear measurement, after performing a point-source aperture-correction on the nuclear flux to the central $3\,\times\,3$ spaxel aperture. The same approach was taken to calculate the extended emission of the 63$\,\micron$ dust continuum, which is also used to scale the {\it IRAS} FIR flux and \LIR\, down to the apertures used to extract the PACS spectra (see DS13 for more details). Errors in the line and continuum fluxes have been propagated into the analysis of, e.g., \CIIno/FIR. To calculate the IR luminosity surface density (\SIR) of the extended emission we use the projected physical area covered by the $3\,\times\,3$ PACS spaxel box centered around the galaxy minus that of the nuclear spaxel. For the starburst nuclei, however, we use the beam-corrected MIR sizes estimated from the Spitzer/IRS spectra and MIPS imaging, $\SIR\,=\LIR\,/\pi r_{\rm MIR}^2$ (see DS13).
% This allowed us to measure the \textit{intrinsic} (instrumental-beam corrected) nuclear starburst size of each galaxy at scales up to $\sim\,10$ times smaller than the size of a PACS spaxel.
%, avoiding the need to assume a filling-factor.

A galaxy is considered as spatially resolved when
% the fraction of extended emission (FEE) of the \CIIno\, line is $\geq\,$0.3, that is, when 
more than 30\% of the \CIIno\, and FIR continuum flux is extended, as defined above.
%We also impose a similar constraint to the continuum emission measured at 158 and 63$\mu$m to ensure we do not systematically select sources with faint extended continuum (and thus FIR). 
This threshold is robust against observations of unresolved sources in which mis-pointings of up to $\sim\,3\arcsec$ would introduce a systematic underestimation of the nuclear flux that would mimic extended emission\footnote{ We have used the new task \href{http://herschel.esac.esa.int/hcss-doc-12.0/load/pacs\_urm/html/herschel.pacs.spg.spec.SpecExtendedToPointCorrectionTask.html}{"specExtendedToPointCorrection"} available in HIPEv12 to estimate that a ~3\arcsec\, mis-pointing within the central spaxel would increase artificially the fraction of extended emission only up to 23\%.}. Because of possible contamination from a nearby companion galaxy, we exclude from this study LIRG systems in which individual galaxies are closer than 23.5$\arcsec$ from each other (1/2 of the PACS FoV). A total of 60 galaxies (25\% of the sample) show extended \CIIno\, and FIR emission meeting these constraints. Their average extended \CIIno\, emission is 52\%, with a maximum of 75\%. The projected radial distance at which the extended emission is detected ranges from 1 to $\sim\,$12.5\,kpc for the closest and farthest resolved galaxy, respectively, with a median of 4.3\,kpc.
%As explained in \cite{DS2010b}, there is a weak inverse dependence of the FEE (in that case of the MIR continuum) with the distance to the source, and therefore with the \LIR\, of galaxies, as more luminous systems tend to be located farther away. In that work we found that galaxies with \LIR\,$\gtrsim\,10^{11.8}\,\Lsun\,$ present a significant drop of their FEE. However, we also showed that this is not associated with a distance effect. As for the rest of the sample, a K-S test comparing the distance and luminosity distributions of the GOALS galaxies with \LIR\,$<\,10^{11.8}\,\Lsun\,$ and those of galaxies classified as extended in this work, confirms that both sub-samples are indistinguishable in both quantities (D-values of 0.06 and 0.14 and $p$-values of 0.99 and 0.34, respectively).
While more luminous GOALS systems tend to be located farther away, a K-S test comparing the distance and luminosity distributions of the GOALS galaxies with \LIR\,$\leq\,10^{12}\,\Lsun\,$ and those of galaxies classified as extended in this work, confirms that both sub-samples are indistinguishable in both quantities (D-values of 0.20 and 0.15 and $p$-values of 0.04 and 0.22, respectively).

%We have to note, though, the implicit assumption that we consider here, which is that the star formation responsible for the extended \CIIno\, emission, FIR flux, and \LIR\, fills-in completely the physical area enclosed by the extraction aperture (see Section~\ref{ss:pacsobs}). Of course this is an overestimation, since a filling-factor of unity is the most extreme upper limit to the actual area covered by active star-forming regions within the disk of a galaxy. On the other hand, the physical sizes obtained for the nuclei of our LIRGs were derived from \textit{Spitzer} IRS or MIPS data at angular resolutions of $\sim\,3-6\arcsec$ (\citealt{DS2010b}).
%(after accounting for the unresolved PSF component)

\section{Results}\label{s:results}

\subsection{\CIIno\, Deficit and FIR colors of Extra-nuclear Star Formation}\label{ss:extdeffir}

\begin{figure}[t!]
\vspace{.25cm}
\epsscale{1.15}
\plotone{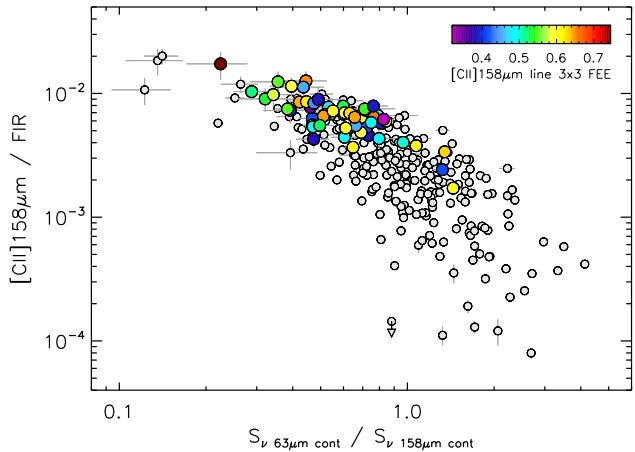}
\vspace{.25cm}
\caption{\footnotesize The ratio of \CII\, to FIR flux as a function of the $S_\nu\,$63\,$\mu$m/$S_\nu\,$158\,$\mu$m continuum flux density ratio for the extended (colored circles) and nuclear regions (gray circles; taken from DS13) of spatially-resolved galaxies in the GOALS sample. Circles of different colors (see color bar) indicate the fraction of \CIIno\, emission that is extended (FEE) outside of the aperture-corrected central PACS spaxel containing the nuclear emission (see section~\ref{ss:pacsobs}). (A color version of this figure is available in the online journal.)}\label{f:ciifirvss63s100}
\vspace{.5cm}
\end{figure}

Figure~\ref{f:ciifirvss63s100} shows the \CIIno/FIR\, ratio as a function of the $S_\nu\,$63\,$\mu$m/$S_\nu\,$158\,$\mu$m continuum flux density ratio (63/158\,$\mu$m hereafter)\footnote{ For comparisons with {\it IRAS}-derived FIR colors, an equation to convert PACS 63/158\,$\mu$m into \textit{IRAS} 60/100$\,\mu$m can be found in the Appendix of DS13.} for the extra-nuclear emission observed in spatially resolved LIRGs in the GOALS sample. As expected, the extended emission regions follow the trend found for the LIRG nuclei (DS13) suggesting that the \CIIno\, deficit is related to the average temperature of the dust (\Tdust). However, the extended regions show smaller \CIIno\, deficits (stronger \CIIno\, to FIR emission) and lower 63/158\,$\mu$m ratios than the majority of the nuclei.
%However, they mostly display \CIIno\, deficits and FIR colors similar to those seen in the nuclei with the coldest dust temperatures.
%In fact, except for three sources, all extra-nuclear regions have a \CIIno/FIR$\,\geq\,4\,\times\,10^{-3}$. This is similar or larger than the mean ratio found in the center of pure star-forming LIRGs (DS13). 
The extended emission has a mean and median \CIIno/FIR\, ratio of $(6.9\,\pm2.6)\,\times\,10^{-3}$ and $6.6\,\times\,10^{-3}$, respectively, which is a factor 2 larger than the LIRG nuclei. Figure~\ref{f:ciifirvss63s100} also reveals that the fractional extent of the \CIIno\, emission does not correlate with the FIR color of the extra-nuclear regions suggesting that galaxies with larger extended emission fractions do not have systematically colder or warmer \Tdust.
%, as galaxies with different \CIIno\, FEEs are randomly located in the parameter space. Instead, 
In the majority of LIRGs the extended emission has 63/158\,$\mu$m$\,\lesssim\,$1 and \CIIno/FIR\,$\geq\,4\,\times\,10^{-3}$, similar to those found in the extended disks of normal star-forming galaxies (\citealt{Madden1993};
%\citealt{Lord1996b}; \citealt{Braine1999}; 
\citealt{Nikola2001};
%\citealt{RF2006};
\citealt{Mookerjea2011}; \citealt{Croxall2012}), as well as in the diffuse component of the ISM in the Galactic plane (\citealt{Shibai1991}; \citealt{Bennett1994}; \citealt{Nakagawa1998}; \citealt{Yasuda2008}). This is consistent with our previous findings based on \textit{Spitzer}/IRS spectroscopy showing that the physical properties of the extra-nuclear, kpc-scale star formation in LIRGs are very similar to those of normal star-forming galaxies with lower IR luminosities ($\LIR\,\sim\,10^{10-11}\,\Lsun$), and that the diversity of their integrated MIR spectra is driven by the nuclear, few central kpc starburst (\citealt{DS2011}).

\begin{figure}[t!]
\vspace{.25cm}
\epsscale{1.15}
%\plotone{/data1/goals/herschel/pacs/analysis/extended-fig2a-1.ps}
%\vspace{.25cm}
\plotone{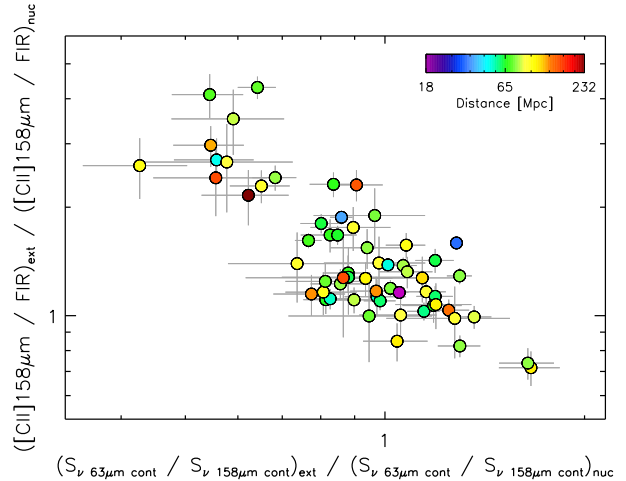}
\vspace{.25cm}
\caption{\footnotesize The ratio of \CIIno/FIR\, between the extended and nuclear emission of LIRGs as a function of the extended to nuclear FIR color. Circles of different colors indicate the distance to the galaxy. (A color version of this figure is available in the online journal.)}\label{f:ciifirvsnucs63s100}
\vspace{.5cm}
\end{figure}

The differences between the nuclear and extended emission are more clearly seen in Figure~\ref{f:ciifirvsnucs63s100}, which shows that the ratio between the FIR color of the extra-nuclear and nuclear regions is clearly correlated with the excess of \CIIno\, deficit seen in the nuclei with respect to that observed in the disks, regardless of the distance at which the extended emission is measured or on the distance to the galaxy (see color-coding). That is, for a given \Tdust\, of the extended emission, LIRGs with warmer nuclei show larger differences between their nuclear and extra-nuclear \CIIno\, deficits.

\subsection{Extended IR Luminosity Surface Densities}\label{ss:irlsd}

\begin{figure*}[ht!]
\vspace{.25cm}
\epsscale{1.15}
\plottwo{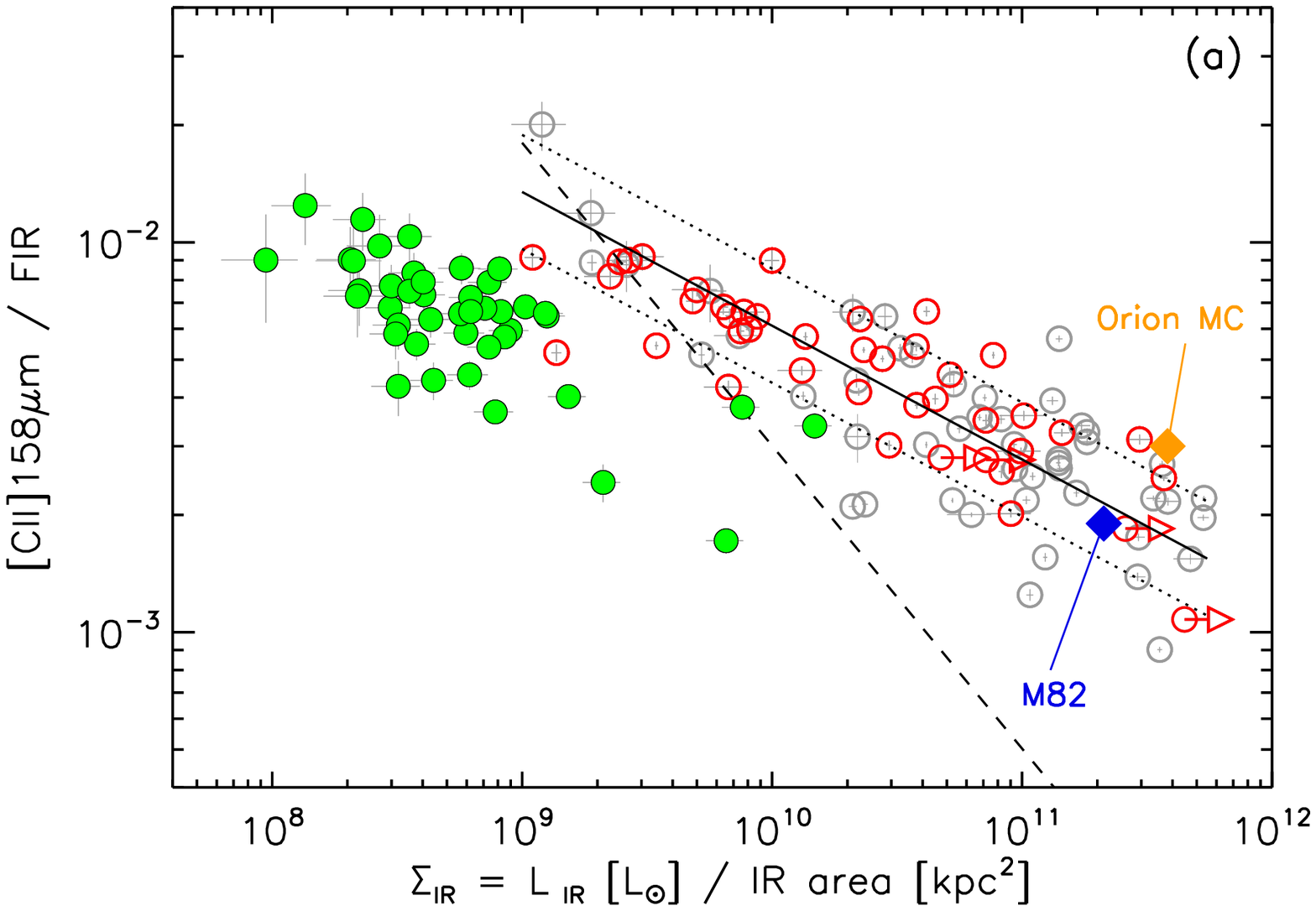}{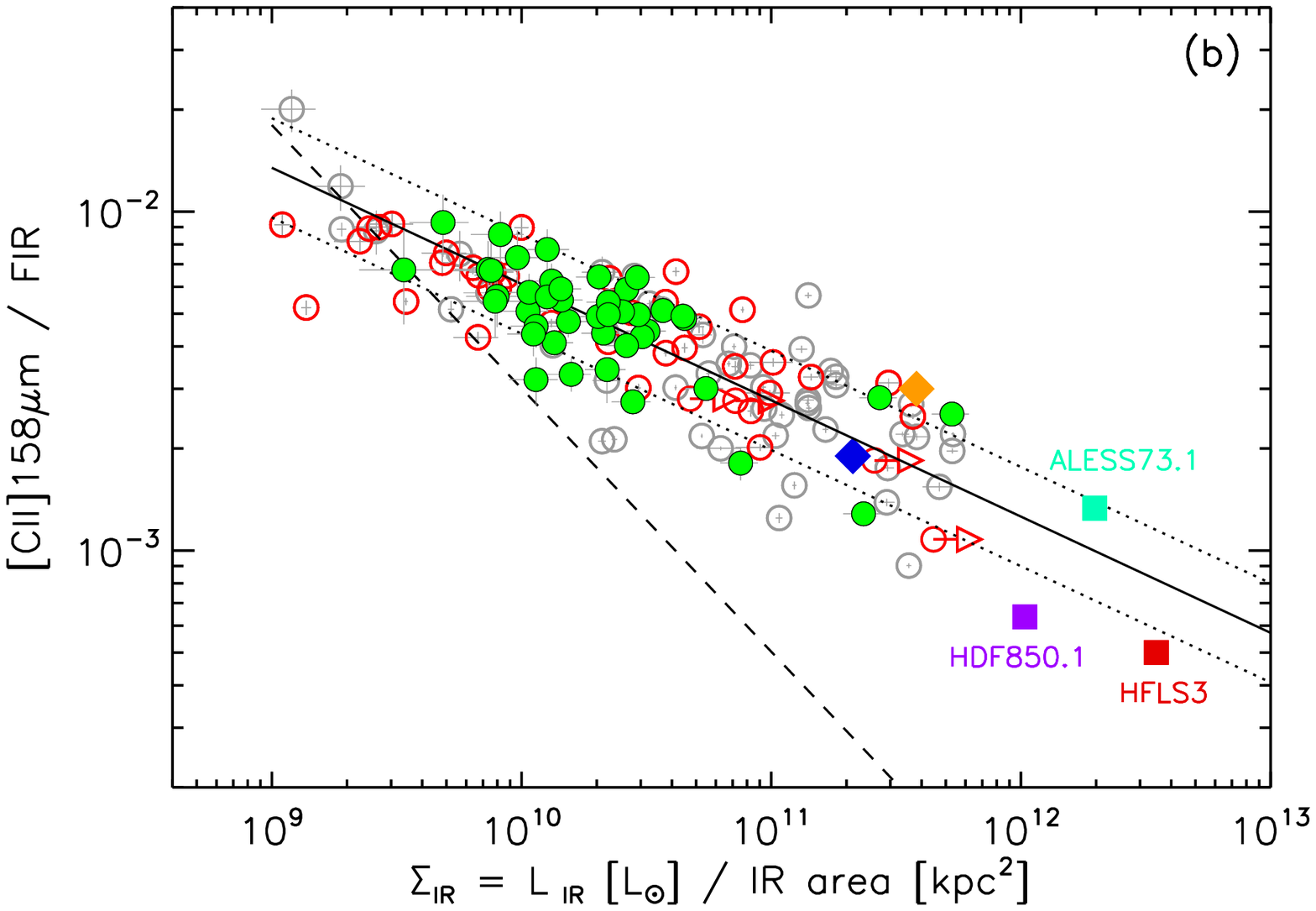}
\vspace{.25cm}
\caption{\footnotesize (a) The ratio of \CII\, to FIR flux as a function of the IR luminosity surface density, \SIR, of the extra-nuclear regions for 43 spatially-resolved star-forming galaxies in the GOALS sample (green circles). Gray open circles represent the LIRG nuclei. Those having an associated measurement of their extended emission are plotted in red. 
% with available nuclear sizes and no significant contribution from an AGN to the nuclear flux (\PAHc\, EW$_{\rm nuc}\,\geq\,0.5\,\mu$m). Each nucleus is connected to its extended emission measurement with a randomly-colored line. The IR area of the extened emission is the physical area covered by a $3\,\times\,3$ PACS spaxel box centered around the galaxy minus the nuclear spaxel (see Section~\ref{ss:pacsobs}). The IR area of the nuclear emission is derived from the MIR size of the starburst region as in DS13 ($\SIR\,=\LIR\,/\pi r_{\rm MIR}^2$).
The definitions of the IR area used for the nuclear and extended regions are given in section~\ref{ss:pacsobs}.
The location of the Orion molecular cloud is shown as an orange diamond (\citealt{Stacey1993}), while that of the starburst in M82 is shown as a blue diamond.
%(\citealt{Rieke1980}; \citealt{Joy1987}; \citealt{Lord1996a})
The black solid line represents a fit to the LIRG nuclear data. The dotted lines are the 1$\sigma$ uncertainties. (b) Same as in (a) but we have increased the \SIR\, of the extended regions by a factor of 1/0.06, equivalent to the difference in beam filling-factor between the extra-nuclear regions and the starburst nuclei (see text). The black dashed line is the expected relation between \CIIno/FIR\, and \SIR\, derived from the PDR models of \cite{Wolfire1990} when $\G0\,\propto\,R^{-2}$.
% for a geometrically thick star-forming region in which a central source heats the surrounding molecular clouds, and assuming $\CIIno\,=\CIIno_{\rm PDR}/0.7$. 
The red square is the nearly-maximal starburst galaxy HFLS~3, at $z\,\sim\,6.34$ (\citealt{Riechers2013}) and the purple and cyan squares are HDF~850.1 and ALESS~73.1, two sub-millimeter galaxies at $z\,\sim\,5.19$ (\citealt{Neri2014}) and 4.76 (\citealt{DeBreuck2014}). (A color version of this figure is available in the online journal.)
%, both very close to the extrapolation of the correlation found for the LIRG nuclei.
}\label{f:ciifirvsigmaIR}
\vspace{1.cm}
\end{figure*}

As shown in DS13, there is a clear trend for pure star-forming LIRGs having more compact IR emission to display lower \CIIno/FIR\, ratios.
%, going from $10^{-2}$ to $\sim\,10^{-3}$.
% (see Figure~\ref{f:ciifirvsigmaIR} (a), gray circles). 
%This implies that the \CIIno\, deficit observed in the nuclei of LIRGs is not caused by a rise of AGN activity\footnote{At least in this regime, but it could be possible in the most extreme cases, at \CIIno/FIR\,$<\,10^{-3}$, where the AGN can contribute significantly to the \LFIR.}, but instead is a fundamental property of the starburst itself.
%The results obtained in Section\ref{ss:extdeffir} already suggest that this is the case, as we find that the \CIIno/FIR\, ratio in extended emission found in our LIRGs is rather constant, with typical values from ranging between $6-10\,\times\,10^{-3}$.
%If this is true, a similar trend should be also evident among the extra-nuclear star formation found in our LIRGs.
Figure~\ref{f:ciifirvsigmaIR} (a) shows that the extended emission regions in spatially-resolved galaxies follow an anti-correlation between \CIIno/FIR\, and \SIR\, similar to that seen for the LIRG nuclei, with slopes of $-0.39\pm 0.07$ and $-0.35\pm 0.03$ respectively, suggesting that these correlations are likely driven by the same physical process.
However, the extra-nuclear regions (green filled circles) are offset by more than an order of magnitude towards lower \SIR\, and show higher \CIIno/FIR\, ratios on average than their nuclei (red open circles).

To understand this offset we can decompose it as a simultaneous shift in \CIIno/FIR\, and \SIR\, along the slope of the observed trend, plus an horizontal shift in \SIR.
%This basically aligns the extended emission regions with their nuclei. 
The shift along the correlation is determined by the ratio between the median \CIIno/FIR\, measured in the extra-nuclear regions and that of the starburst nuclei.
The additional horizontal shift in \SIR\, represents the beam filling-factor of the star-forming regions in the LIRG disks with respect to that of the nuclei. We estimate this relative filling-factor to be 0.06$^{+0.13}_{-0.04}$ by minimizing the difference between the median \SIR\, of the extended and nuclear emissions (see Figure~\ref{f:ciifirvsigmaIR} (b)).
%, once the correction along the trend is made.
% The last step is equivalent to maximizing the occupied area to a level equivalent to what is seen in the starburst region of galaxies showing similar \CIIno/FIR\, ratios, aligning the extra-nuclear and nuclear points on a common correlation.
The uncertainties in the filling factor account for the dispersion of the \CIIno/FIR\, vs. \SIR\, correlation in the x-axis as well as for variations in the properties of the ISM that would modify the \CIIno/FIR\, ratio,
% down to the level of the nuclear regions without affecting the \SIR, 
such as changes in the fractional contribution of \CIIno\, emission associated to the diffuse ionized medium.
%apply a correction of 0.08 to the area used to calculate the extended \SIR, we obtain Figure~\ref{f:ciifirvsigmaIR} (right). The area scaled in this way pushes the extended emission regions right on the correlation found for the nuclei.
%Summarizing, under the assumptions described above, we find that the beam filling-factor of the extra-nuclear star-forming complexes responsible for the \CIIno\, and FIR flux in the LIRG disks is $\sim\,$6\% of the filling-factor for the starburst nuclei.
%(\textbf{Any idea if this number seems reasonable?}).
% Assuming that the \CIIno\, emission contributed by the diffuse ($\ne\,\sim\,10\,\c-mmm$), ionized ISM is small such that most of the \CIIno\, arises from PDRs or \HII\, regions, this filling-factor is equivalent to an average distance between \CIIno-emitting knots of star formation in a LIRG disk $\sim\,6$ times their average size. Evidently, the precise value depends on the distance to the center of the galaxy, since it is known that the light profile of galactic disks (as well as the number of star-forming regions) decays exponentially with galacto-centric distance. Indeed, part of the scatter seen in the $x$-axis of the correlation may be produced by this, as the PACS FoV covers a different physical area over the galaxies depending on the distance at which they are located.

This methodology has been recently used in combination with the PDR models developed by \cite{Wolfire1990} to argue for the presence of large-scale star-formation in high-redshift galaxies (\citealt{HD2010}; \citealt{Stacey2010}; \citealt{Ferkinhoff2014}). These models relate the radius ($R$) and \LIR\, of the starburst region to the strength of the UV inter-stellar radiation field (ISRF), measured in units of the local Galactic value (\G0\,=\,1.6\,$\times\,10^{-3}$\,erg\,s$^{-1}$\,cm$^{-2}$; \citealt{Habing1968}). We can compare the relative beam filling-factor of the extra-nuclear star-forming regions derived from Figure~\ref{f:ciifirvsigmaIR} to that predicted by the models. \cite{Wolfire1990} consider two geometric configurations for the starburst region. One in which the molecular clouds are immerse in a smooth ISRF and another in which a central, ionizing source is surrounded by a geometrically-thick, molecular medium. Each one follows a relation such that $\G0\,\propto\,\lambda\,\LIR\,/\,R^3$, and $\G0\,\propto\,\LIR\,/\,R^2$, respectively, with $\lambda\,$ being the mean free path of a UV photon within the starburst region. If we assume a hydrogen density, \nH, in the range of $\sim\,10^{2-5}\,\c-mmm$, we can use Figure~4 of \cite{Stacey2010}, which is based on the PDR models from \cite{Kaufman1999}, to estimate the \CIIno/FIR\, ratio as a function of \G0. Although neither model fits our observed trend of \CIIno/FIR\, with \SIR, the model with a compact, nearby ionizing source, i.e., $\G0\,\propto\,R^{-2}$ (dashed line in Figure~\ref{f:ciifirvsigmaIR}), is closest to our data. If we use M82 as our reference starburst (with the following properties: $\G0\,=\,1000\,$ and $\LIR\,\sim\,3\,\times\,10^{10}\,\Lsun$, \citealt{Lord1996a}; $D\,\sim\,420\,$pc, \citealt{Joy1987}), the beam filling factor predicted by this configuration for the extra-nuclear star-forming regions would be $\sim\,1$\% that of M82. This value is significantly lower than our inferred 6\%, when we compare them to the LIRG nuclei using the observed correlation in Figure~\ref{f:ciifirvsigmaIR}. We explore the reasons for this discrepancy below.

\section{Discussion}\label{s:summary}

The difference between the filling factor derived from our data and that estimated using the PDR models is due to the slope of the trend found for the LIRG nuclei, which is still significantly shallower than that implied by the PDR model.
%, and the fact that the template we use for its derivation in each case (M82 and the LIRG nuclei) have different \CIIno/FIR\, ratios and \SIR. 
Using the relationship between \CIIno/FIR\, and \G0\, in (\citealt{Stacey2010})\footnote{See also the Photo Dissociation Region Toolbox (PDRT) models: http://dustem.astro.umd.edu/pdrt/ (\citealt{Tielens1985}; \citealt{Wolfire1990}; \citealt{Hollenbach1991}; \citealt{Kaufman1999}).}, a spatial distribution of the star formation of $\G0\,\propto\,\LIR\,/\,R^2$ is equivalent to log(\CIIno/FIR)\,$\propto\,-0.78\,\times\,$log($\LIR/R^2$) (see dashed line in Figure~\ref{f:ciifirvsigmaIR}). This significantly differs from the slope fitted to the data, $-0.35$. The difference in the slopes may arise at least in part from the fact that as the radiation field becomes more intense, not only the gas heating efficiency is reduced due to the increased ionization of small dust particles but dust grains could also be intercepting a larger fraction of the UV radiation before it reaches the PDR (see \citealt{Abel2009}). Since this dust will be warmer, the natural consequence is also a smaller size for the starburst, as measured in the MIR, and hence a larger \SIR.
%This also has an effect of raising the dust temperature and making the starburst appear more compact in the MIR, thus increasing the \SIR.
In fact, the \CIIno/FIR\, ratio is correlated with the depth of the 9.7$\,\micron$ silicate absorption feature (DS13), which measures the obscuration towards the warm dust continuum originating from the nuclear star-forming regions. It is therefore a combination of a decreased heating efficiency plus warm dust inside the PDR and likely within the \HII\, regions, that causes a drop in \CIIno/FIR\, and a rise in \SIR. In these dusty sources, other FIR emission lines arising from the dense ionized gas (e.g., \NIIno, \NIIIno, or \OIIIno) also show deficits with respect to the FIR emission (\citealt{Gracia-Carpio2011}), suggesting that a significant amount of dust may indeed be present inside the \HII\, regions.
The PDR model presented in Figure~\ref{f:ciifirvsigmaIR} not only has a steeper slope than the trends observed for the nuclei and extended regions of local LIRG, but the majority of the nuclear data, mostly above $\SIR\,\sim\,10^{10}\,\Lsun\,\k-pcc$, are located to the right of the model prediction. Since the model assumes a constant filling-factor of unity for the starburst region, this implies a beam filling-factor for the LIRG nuclei increasingly larger than unity. A possible interpretation is that, at a given \CIIno/FIR\, ratio, we can consider the trend with \SIR\, found for the LIRG nuclei as "maximal", i.e., where the nuclear starbursts have the highest --most compact-- \SIR\, possible, with multiple star-forming regions overlapping along the line of sight within the beam.

These findings may be useful for studies of IR-luminous galaxies detected in high-redshift surveys for which it is not possible to resolve the \CIIno\, or IR emission.
%, or for which a size of the starburst region cannot be determined.
%Given a \LIR\, and an integrated \CIIno/FIR\,$\,\lesssim\,4\,\times\,10^{-3}$ (i.e., a galaxy in the "starbursting regime"), it is possible to predict the effective size of its starburst region. 
As an example, in Figure~\ref{f:ciifirvsigmaIR} (b) we show HFLS~3 (\citealt{Riechers2013}), a near-maximal starburst galaxy at $z\,\sim\,6.34$, and two sub-millimeter galaxies, HDF~850.1 (\citealt{Neri2014}) and ALESS~73.1 (\citealt{DeBreuck2014}), at $z\,\sim\,5.19$ and 4.76, respectively. All galaxies have a \CIIno\, deficit and \SIR\, broadly consistent with the extrapolation of the correlation derived from the nuclei of local LIRGs (DS13).
We note that the \SIR\, of the high-$z$ sources may actually be underestimated because their sizes are measured in the rest-frame FIR continuum, which is dominated by the cold dust. 
% We note that the projected sizes used to calculate the \SIR\, of these galaxies are those of the cold dust emission (at $\sim\,160\,\mu$m), making the \SIR\, a lower limit in our plot since the LIRG nuclear sizes are measured in the MIR. 
%If HFLS~3 and HDF~850.1 follow the trend found for the LIRG nuclei,
%is truly a high-$z$ analog of local star-bursting 
%(i.e., they are galaxies out of the "main-sequence"),
%LIRGs and ULIRGs with no significant AGN contribution to the FIR emission, based on a simple extrapolation of our trend, we estimate that the difference between the size of its (warm) MIR-emitting and (cold) FIR-emitting dust is a factor of $\sim\,1.5-2.5$. 
%the size of their starbursts should be about $\sim\,1.5-3$ smaller in radius than seen in the FIR. 
But if the reported sizes are correct, the offset of HFLS~3 and HDF~850.1 may imply beam filling-factors for the star-forming complexes $2-5$ times larger than local LIRG disks would have with similar \CIIno/FIR\, --- or, alternatively, these high-$z$ sources would be showing properties similar to those of local ULIRG nuclei in GOALS, but extended over larger physical scales.

\section*{Acknowledgments}

We thank the anonymous referee for her/his useful comments and suggestions. We would also like to thank Mark Wolfire for providing us with very useful insights regarding the PDR models. V.C. would like to acknowledge partial support from the EU FP7 Grant PIRSES-GA-2012-316788. This work is based on observations made with the \textit{Herschel} Space Observatory, an European Space Agency Cornerstone Mission with science instruments provided by European-led Principal Investigator consortia and significant participation from NASA. The \textit{Spitzer} Space Telescope is operated by the Jet Propulsion Laboratory, California Institute of Technology, under NASA contract 1407. This research has made use of the NASA/IPAC Extragalactic Database (NED), which is operated by the Jet Propulsion Laboratory, California Institute of Technology, under contract with the National Aeronautics and Space Administration, and of NASA’s Astrophysics Data System (ADS) abstract service.\\

%\bibliographystyle{/Users/tanio/mypapers/apj}%este estilo nombra en la lista de referencia solo a los tres primeros  autores
%\bibliography{/Users/tanio/mypapers/bib}{}

\end{document}